\documentclass{article}
\usepackage[utf8,ansinew]{inputenc}
\usepackage[german,english]{babel}
\usepackage[ansinew]{inputenc}
\usepackage{amsmath}
\usepackage{mathrsfs}
\usepackage{amsbsy}
\usepackage{tabularx}
\usepackage{color}
\usepackage[nottoc,numbib]{tocbibind}
\usepackage[round]{natbib}
\usepackage{fancybox} 
\usepackage[arrow, matrix, curve]{xy}
\usepackage{graphicx} 
\usepackage[switch,columnwise]{lineno}
\usepackage{endnotes} % Fuer Fussnote
\usepackage{framed}
\usepackage{tikz}
\usetikzlibrary{shapes,angles,quotes}
\graphicspath{{./Abbildungen/}} 
\usepackage{mathrsfs}
\DeclareMathAlphabet{\mathscrbf}{OMS}{mdugm}{b}{n}

\usepackage[normalem]{ulem} % zum Text rausstreichen

\let\oldequation\equation
\let\oldendequation\endequation

\renewenvironment{equation}
  {\linenomathNonumbers\oldequation}
  {\oldendequation\endlinenomath}

\usepackage{enumitem}

\usepackage{fancybox} 
\usepackage[T1]{fontenc} 
\usepackage[arrow, matrix, curve]{xy}
\usepackage{amsthm}
\usepackage{dsfont}
\usepackage{graphicx} 

\usepackage{tabularx}
\usepackage{amssymb} %fÃ¼r doppelte Buchstaben, wie die Reellen Zahlen.
\usepackage{amsmath} %zB fÃ¼r matrizen
\usepackage{color}
\usepackage{framed}
\usepackage{endnotes} % Fuer Fussnote

\usepackage[onehalfspacing]{setspace}
\usepackage[round]{natbib}
\usepackage{tikz}
\usetikzlibrary{shapes,angles,quotes}

\usetikzlibrary{calc}
\usetikzlibrary{matrix}
\usepackage{pgfplots}
\pgfplotsset{compat=1.11}
\usepackage{braket}

\theoremstyle{definition}

\usepackage{varioref}
\usepackage{graphicx} %Loading the package

\newcommand{\DSI}{{\rm DSI}}

\newcommand{\be}[1]{\begin{equation}\label{#1}}
\newcommand{\ee}{\end{equation}}

\newcommand{\beq}{\begin{eqnarray}}
\newcommand{\eeq}{\end{eqnarray}}
\newcommand{\beqs}{\begin{eqnarray*}}
\newcommand{\eeqs}{\end{eqnarray*}}
\newcommand{\bequ}{\begin{equation}}
\newcommand{\eequ}{\end{equation}}

\usepackage[% fast immer als letztes Paket
	colorlinks,% farbige Links statt Rahmen
	linktocpage,% verlinke Seitenzahl, satt Titel
	linkcolor=black,% Linkfarbe(Text)
	urlcolor=black,% URLfarbe (WEB)
	citecolor=black,%
]{hyperref}

\title{Reducing the Dynamic State Index to its main information using Principal Component Analysis}
\author{Annette M\"uller, Nikolai Spaeth, Johanna Bollow, Iris Eder, \\ Alina Fischer,
Jana Frenzel, Annika G\"ornt, Kunyan Hao,\\  Johanne Ilchmann, Isabel Kasner, Corinna Langwald, \\ 
Steffen Laube, Flavio Maggioni, Beatrice Neiszt-Pllana, \\
Daniel A. Redant, Lisa Rogge, Sus Sama, David Scheer, \\ Inga Spreen,
Patrick W. Varchmin, Vanessa R. Zabel, \\
Wei Zhang, Anika Zibell}
\begin{document}

\maketitle
\section*{Abstract}

The Dynamic State Index is a scalar quantity designed to identify atmospheric developments such as fronts, hurricanes or specific weather pattern. The DSI is defined as Jacobian-determinant of three constitutive quantities that characterize three-dimensional fluid flows: the Bernoulli stream function, the potential vorticity (PV) and the potential temperature. Here, we tackle the questions (i) if the mathematical formulation of the DSI can be reduced, while keeping the main information, and (ii) does the reduction of the DSI depend on the spatial scale? Applying principle component analysis we find that three of six DSI terms that sum up to the Jacobi-determinant are sufficient for future DSI calculations. 

\section{Introduction}

Describing atmospheric flow fields accurately is a great challenge. Physically, fluid flows are characterized by an interaction of thermodynamic, dynamic and vortex dynamic processes. 
One scalar quantity that couples information of all three processes into one scalar is the so-called Dynamic State Index (DSI) introduced by \citet{Nevir2004}.
\citet{Schaer1993} introduced the steady wind that characterizes the basic state of a fluid.
The DSI is designed to be zero for the adiabatic, inviscid, steady basic state and non-zero DSI values localize deviations from the basic state, i.e. diabatic, viscous and non-steady processes.  
For example, consider a persistent high pressure area in the midlatitudes. This region  would be indicated by vanishing DSI values, whereas the jet stream surrounding the high pressure area would be signalized by non-zero DSI values \citep{mueller2019using}. In particular, the DSI signal indicates e.g. regions of precipitation as shown in \citet{Claussnitzer2008} or \citet{Mueller2018}. \citet{Weber2008} highlights the importance of its dipole structure and shows that the DSI can be applied to diagnose hurricanes \citep{Weber2008}. 
In order to detect scale-dependent processes more precisely, the DSI has further been developed for the quasi-geostrophic model and the Rossby model \citep{Mueller2018,mueller2019using}. 
Recently, \citet{Hittmeir2021} introduced DSI variants for moist processes that will be applied to identify and differ between specific  moist processes in future studies. 

Here, we consider the original DSI and search for the parts of the DSI that contribute most to the DSI signal in order to obtain a reduced DSI variant that keeps the main information. Mathematically, the DSI is defined as Jacobian-determinant of the potential temperature $\theta$, the Bernoulli stream function $B$ that contains the kinetic energy and the potential vorticity (short known as PV and denoted as $\Pi$) that captures vortex motion.
The potential temperature is defined as:
\begin{equation}
\theta = T \left( \frac{p_n}{p} \right)^{R/cp},
\end{equation}
where $T$ denotes the temperature, $p$ is the pressure, $c_p$ is the mass specific heat constant for dry air at constant pressure and $R$ is the specific gas constant for dry air.
The Bernoulli function is defined as:
\begin{equation}
     B = \frac{1}{2}\mathbf{v}^2+c_pT+\phi \, 
\end{equation}
with the geoptential $\phi$ and the 3D wind vector $\mathbf{v}$.
The PV is given by:
\begin{equation}
    \Pi = \rho^{-1} \boldsymbol{\xi}_a \cdot \nabla \theta \, ,
\end{equation}
where $\boldsymbol{\xi}_a$ denotes the absolute vortex vector.
Then, the DSI is given by the Jacobian-determinant of these three quantities coupling all the physical information into one scalar: 
\begin{equation} \label{eq:DSIgen1}
\mathrm{DSI}_{\rm PE}:= \Big\vert \frac{\partial(\Theta,B,\Pi)}{\partial(a,b,c)}\Big\vert = \frac{1}{\rho}\Big\vert\frac{\partial(\Theta,B,\Pi)}{\partial(x,y,z)}\Big\vert
= \frac{1}{\rho} \left(\nabla\theta \times \nabla B \right) \cdot \nabla \Pi \, .
\end{equation}
 \citep{Nevir2004}.
Here, $dm = da \ db \ dc = \rho  dx \ dy \ dz$ are the Lagrangian mass coordinates and $\rho$ denotes the density. 
The three quantities in the Jacobi-determinant.  We note the important property that the advection of these three constitutive quantities vanishes, see e.g. \citet{Sommer2009}. In this case the DSI is equals to zero. 

 %The results are in the summer term 2020 at the Freie Universit\"at.

 To calculate this original DSI nine different partial gradients have to be calculated:  %\textcolor{red}{hinter (3.89)}
\begin{equation} \label{eq:DSIgrads}
\frac{\partial \theta}{\partial x},\frac{\partial \theta}{\partial y},\frac{\partial \theta}{\partial z},\frac{\partial B}{\partial x},\frac{\partial B}{\partial y},\frac{\partial B}{\partial z} ,\frac{\partial \Pi}{\partial x},\frac{\partial \Pi}{\partial y},\frac{\partial \Pi}{\partial z}   \, .
\end{equation}
These gradients form the following six terms:
\begin{equation} \label{eq:terms1-3} 
\text{Term 1} =\frac{\partial \Pi}{\partial p} \cdot \frac{\partial \theta}{\partial y} \cdot \frac{\partial B}{\partial x},
\qquad   
 \text{Term 2} = -\frac{\partial \Pi}{\partial p} \cdot \frac{\partial \theta}{\partial x} \cdot \frac{\partial B}{\partial y },
 \qquad  \text{Term 3} = \frac{\partial B}{\partial p} \cdot \frac{\partial \theta}{\partial  x} \cdot \frac{\partial \Pi}{\partial y} \, .
\end{equation}
\begin{equation}\label{eq:terms4-6} 
\text{Term 4} = -\frac{\partial B}{\partial p} \cdot \frac{\partial \theta}{\partial y} \cdot \frac{\partial \Pi}{\partial x},
\qquad
\text{Term 5} = \frac{\partial \theta}{\partial p} \cdot \frac{\partial B}{\partial y } \cdot \frac{\partial \Pi}{\partial x}, 
\qquad \text{Term 6} = - \frac{\partial \theta}{\partial p} \cdot \frac{\partial B}{\partial x} \cdot \frac{\partial \Pi}{\partial y}\, .
\end{equation}
which, in turn, sum up to the definition of the DSI:
\begin{equation}
DSI = \frac{1}{\rho} \bigg \vert \frac{\partial (\theta, B , \Pi ) }{\partial (x,y,z)} \bigg \vert = \frac{1}{\rho}( \text{Term 1}+\text{Term 2}+\text{Term 3}+\text{Term 4}+\text{Term 5}+ \text{Term 6}) \, .
\end{equation}

The question arises, whether these terms differ in their allocation of information, i.e. which of them contribute most to the DSI signal? Consequently it would be possible to reduce the DSI to the relevant information and thus reduce e.g. the time necessary for computing this index. To solve these questions a Principal Component Analysis will be performed as portrayed in the next sections.

Principal Component Analysis (PCA) can be seen as a multivariate statistical technique to analyze data. It is designed to reduce the dimension of a data set, i.e. the size of the data set, while retaining the most important. 
For this purpose the data set is transformed to a new set of uncorrelated variables, the so-called principle components (PCs), which are ordered in a descending manner with regard to their information content respectively their ability to describe variance present in all of the original variables \citep{jolliffe2016principal}. 
Hence, the PCs represent identified orthogonal directions (eigenvectors), along which the variation in the data is maximal. 
The new set of extracted variables then enable to visually assess patterns of similarity and differences between samples \citep{ringner2008principal,abdi2010principal}. To sum up, the PCA can be used to structure, simplify and exemplify a complex data set by approximating multiple statistical variables through a few significant linear combinations \citep{bailey2012principal}. 

In the following, we will apply PCA to the DSI using the ERA-Interim data set and show, which of the DSI terms provide the most information. 

 \section{Data}
\label{sec:data}

For the analysis the ERA5 reanalysis data set on pressure levels of the ECMWF is used \citep{HersbachERA}. It has a lat-lon grid of $0.25^{\circ} \times 0.25^{\circ}$. Vertically, the DSI is calculated on the pressure levels 450, 500, 550, 600, 650, 700, 750, 775 h Pa. 
The hourly data of the month July 2021 is considered in the spatial region of about Germany $46^{\circ} N - 56^{\circ} N$, $5^{\circ} E - 16^{\circ} E$.
The month July is considered, because of the high number of extreme precipitation events. One extreme event took place at July 14th, where accumulated precipitation of 162.4 $l/m^2$ were measured at Wipperf\"urth-Gardeweg.  
The resolution leads to 10981440 data points for each of the six DSI terms, i.e. in total $10981440 \times 6 = 65888640$ data points.
In order to analyse the weights of the DSI terms, the data is standardized using Pythons \emph{scikit-learn} and \emph{StandardScaler}. For the illustrations of the case study, a larger spatial domain of the Euro-Atlantic is considered.

\section{Applying the PCA to the six DSI terms}

The aim of PCA is to reduce a large data set with a lot of variables, while retaining as much as possible information \citep[see, e.g.][]{jolliffe2016principal}.
In order to fulfill this idea,  the data set is transformed to a new set of variables. These new variables are ordered such that the first few contain the main variation of the data. These variables are called principle components. 
To apply a principle component analysis, short PCA, to the DSI terms we use the pythons modules \emph{pandas, numpy, seaborn, scipy, matplotlib} and \emph{sklaern}.

%\begin{figure}
%    \centering
%    \includegraphics[scale = .38]{Scatterplot_large_scale.png}
%    \caption{Scatterplot of the DSI terms defined in \eqref{eq:terms1-3} and \eqref{eq:terms4-6} }
%    \label{fig:Scatterplot_large}
%\end{figure}
%

\begin{figure}%
    \centering
\includegraphics[scale=.6]{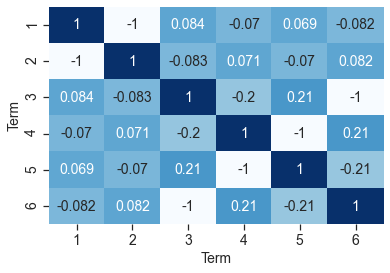}
    \caption{\small{Covariance matrix of the terms \eqref{eq:terms1-3} and \eqref{eq:terms4-6} }}%
    \label{fig:CovMat_large}%
\end{figure}

\begin{figure}%
    \centering
\includegraphics[scale=.3]{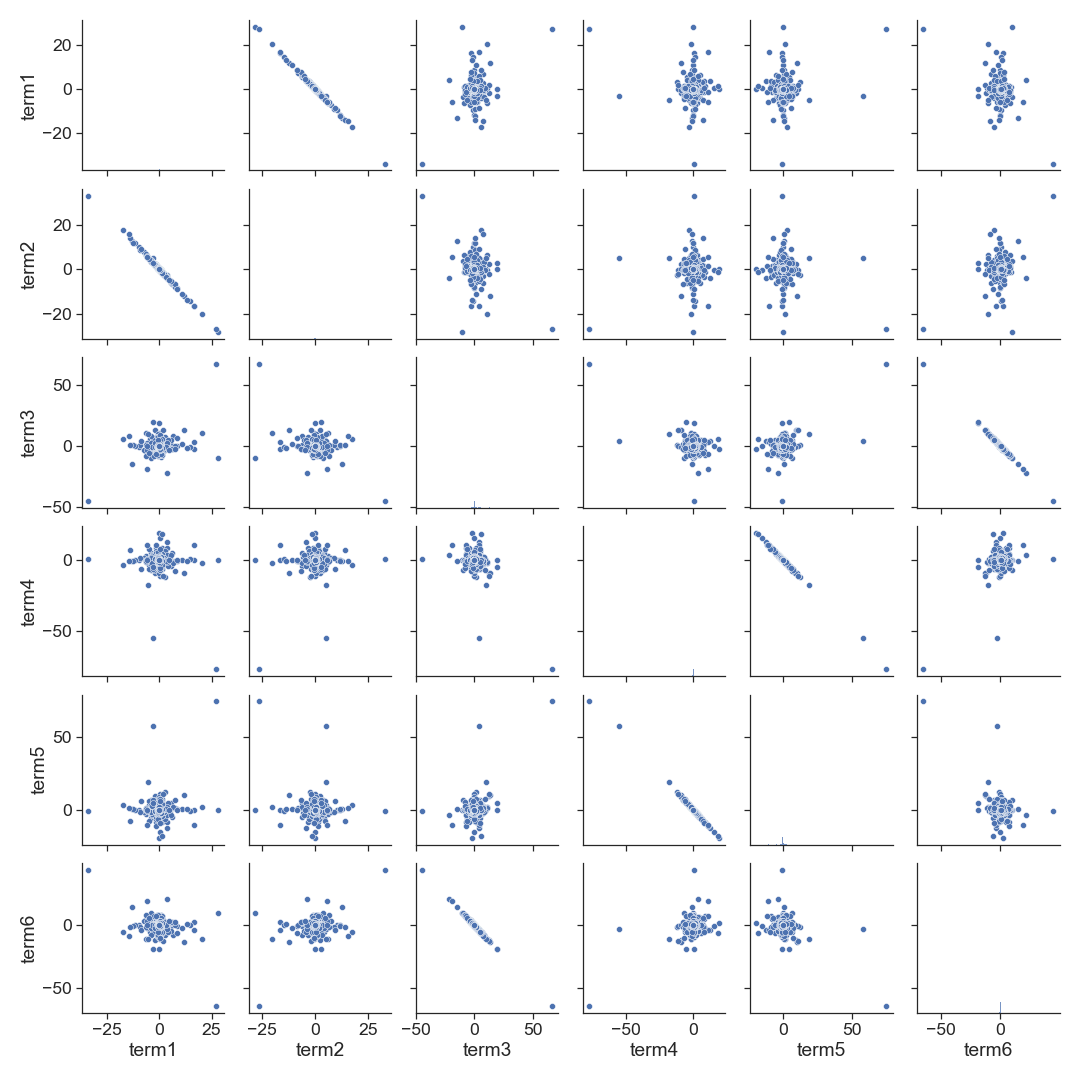}
   \caption{\small{The Scatterplot matrix of the terms \eqref{eq:terms1-3} and \eqref{eq:terms4-6} depicts the results of Fig. \ref{fig:CovMat_large} graphically.}}%
    \label{fig:Scatterplot_PCA_1}%
\end{figure}

Using the ERA5 data described in Se. \ref{sec:data} the six terms of the DSI given in \eqref{eq:terms1-3} and \eqref{eq:terms4-6} are calculated and standardized. Fig. \ref{fig:CovMat_large} shows the covariance matrix of the six terms,
see also the Scatterplotmatrix depicted in 
%the appendix 
Fig. \ref{fig:Scatterplot_PCA_1}. 
A clear strong negative linear relationships between term 1 and term 2, term 3 and term 6 as well as between term 4 and term 5 can be recognized. The remaining relationships show a rather small or no linear character. 
We note that the strength of the correlations of the strongly negatively correlated terms depends on the size of the data set. As larger the data set, as stronger the negative correlation. On the other hand, this means that the weaker correlations become weaker with increasing data. This also means that the covariance matrix of case studies of special weather situations might lead to different results. 
%The eigenvalues and eigenvectors are computed and the outcome is shown as 1D array of complex numbers for the eigenvalues of the covariance matrix and as 2D array of the corresponding eigenvectors. In the present case of zero-values for the imaginary part of the complex eigenvalues the additional function eigen\_vals.real is performed for correction.

%
%In the next step, the linear combinations of the vectors that have maximum variance and that are uncorrelated have to be found. 
Next, the eigenvectors and corresponding eigenvalues of the covariance matrix are calculated.
While the eigenvectors (PCs) determine the direction of the new feature space, their eigenvalues indicate the variances.
%The total variance of the original data set can be calculated by the ratio of the sum of the first eigenvalues to the sum of the variances of all original variables \cite{jolliffe2016principal}. 
%The computation of the PCs here follows the approach of eigendecomposition.

%complex eigen values: [3.24746071+0.j 1.46001275+0.j 1.12766514+0.j 0.0108236 +0.j
% 0.11439902+0.j 0.05905625+0.j]
% Eigenvectors: 
% \begin{equation}
%     \begin{split}
% \mathbf{v}_1 &= ( 0.417, -0.473, -0.307, 0.709,  0.069,  0.004)^T \\
% \mathbf{v}_2 &= (-0.413,  0.481,  0.305, 0.701, -0.049, -0.103)^T \\
% \mathbf{v}_3 &= ( 0.462,  -0.021,  0.492, -0.0467, -0.226, -0.700)^T \\
% \mathbf{v}_4 &= (-0.402, -0.476,   0.281, -0.025, 0.677, -0.271)^T \\
% \mathbf{v}_5 &= (-0.339, -0.559, 0.331, 0.030, -0.640, 0.230)^T \\
% \mathbf{v}_6 &= ( 0.406,  0.060, 0.618, 0.038,  0.273,  0.611)^T \\
%     \end{split}
% \end{equation}
% and the real eigenvalues are given by:

The eigenvectors of the covariance matrix are ordered from the highest to the lowest eigenvalue. The eigenvector with the highest eigenvalue is the first principle component and so on.
As eigenvalues we obtain six eigenvalues, where three eigenvalues are greater or equal to 1:
\begin{equation} \label{eq:eigvals}
\begin{split}
e_1 &= 2.506,\ e_2=1.908 ,\ e_3=1.581 \, .
\end{split}
\end{equation}
%EV: array([2.50608943, 1.90778568, 1.58089667]
According to the Kaiser's rule, which retains only those PCs whose eigenvalues \eqref{eq:eigvals}, i.e. variances, exceed 1 because any PC with a variance less than 1 contains less information than one of the original variables and is therefore not worth retaining. 
Therefore, we consider the first three principle components and the principle components four, five and six can be neglected.
The first three PCs are given in table  \ref{tab:PCs}.

\begin{table}[t!] 
    \centering
\begin{tabular}{|c||c c c|} 
\hline
 & \color{gray}PC 1 &	\color{gray}PC 2 &	\color{gray}PC 3 \\
 \hline
 \hline 
\color{gray} term1 & 	-0.277183 &	0.649616 &	-0.034364  \\
\color{gray}term2 &	0.277427 &	-0.649410 &	0.035908 \\
\color{gray}term3 &	-0.463655 &	-0.169960 &	0.506221 \\
\color{gray}term4 &	0.455237 &	0.220587 &	0.494588 \\
\color{gray}term5 &	-0.456891 &	-0.222172 &	-0.491303 \\
\color{gray}term6 &	0.463937 &	0.171350 &	-0.505246  \\
\hline
\hline
\end{tabular} 
\caption{\label{tab:PCs}\small{The three principle components.}}
\end{table}

%We note that the corresponding eigenvectors that we will choose as PCs will are shown in table \ref{tab:Loadings_large} that we will discuss in the next paragraph. 

%%% SMALL SCALE DATA %%%%%%
% Six eigenvectors were found and therefore six PCs with their corresponding real eigenvalues:
% real eigenvalues $3.43,\ 1.66 0.93,\ 5.55e-04,\
%  8.48e-05,\ 1.66e-04$ their Eigenvectors: 

% \begin{equation}
% \mathbf{v_1} = 
% \begin{pmatrix}
% 0.3134506 ,  -0.63411572 , -0.00579542, -0.07436978,  0.69476401, -0.1066981\end{pmatrix},
% %
% \mathbf{v_2} = 
% \begin{pmatrix}
% [-0.31460878 , 0.63292815 , 0.00553933 , -0.07147753,  0.69589672, -0.10493477]
% \end{pmatrix}
% %
% \mathbf{v_3} = 
% \begin{pmatrix}
% 0.44927524,  0.21891564,  0.49818583, -0.56103698,  0.00766025,  0.43268627
% \end{pmatrix}
% %
% \mathbf{v_4} = 
% \begin{pmatrix}
% -0.44633482, -0.22757236,  0.50107198,  0.42411801 , 0.1276041,   0.54933053
% \end{pmatrix}
% %
% \mathbf{v_5} = 
% \begin{pmatrix}
% 0.44846144,  0.22487593, -0.49621697,  0.43050157 , 0.12899227,  0.54782412
% \end{pmatrix}
% %
% \mathbf{v_6} = 
% \begin{pmatrix}
% -0.44787204, -0.21683355, -0.50442263, -0.5562283 ,  0.00708784,  0.43417911,
% \end{pmatrix}
% \end{equation}

\begin{figure}%
    \centering
\includegraphics[scale =.6]{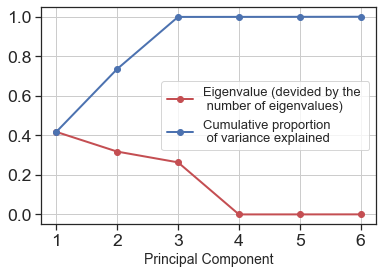}
    \caption{red curve: eigenvalues divided by the sum of eigenvalues, which is equals to six; blue curve: Cumulative proportion of variance explained by the principal components}%
    \label{fig:SreeLarge2}%
\end{figure}

Moreover, the eigenvalues divided by the number of eigenvalues as well as the proportion of variance explained by each PC are shown Fig. \ref{fig:SreeLarge2}.
The depicted normed eigenvalues have the values: $0.418,\ 0.3180,\  0.263$, $0.167364132 10^{-3}$,  $0.271 10^{-3}$, $0.433 10^{-3}$ confirming that the latter three eigenvalues are neglectable small.
Regarding Fig. \ref{fig:SreeLarge2} the variance explained by each PC is divided by the total variance explained by all PCs.
We find that the first PC explains about 41.8\% of the variance of the data, the first and the second PC 73.6\% and the first, second and third PC explain 99.9\%. Therefore the principle components 4-6 are clearly $< .1$\%.
Thus, the first three PCs explain most of the variance confirming that we should use three principle components.

\begin{figure}[t]
    \centering
    \includegraphics[scale=.8]{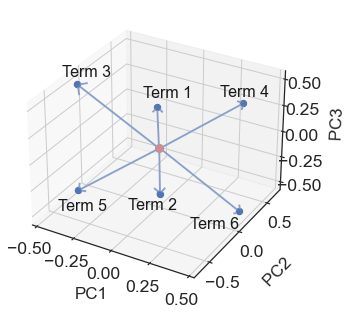}
    \caption{Loading plot, the axis are the principle components, the DSI Terms 1--6 are marked by the blue dots. The red dot marks the common origin of the 6 terms in the feature space.}%
    \label{fig:Loadings}%
\end{figure}
%
% The corresponding loading plot is shown in fig. \ref{fig:Loadings}. 
% In general, one can state that variables with little contribution to a direction have almost zero weight in that loading.
% Variables which have roughly equal influence on defining a direction are correlated with each
% other and will have roughly equal numeric weights.
% Strongly correlated variables, will have approximately the same weight value when they are
% positively correlated. In a loadings plot of e.g. p1 vs p2 they will appear near each other,
% while negatively correlated variables will appear diagonally opposite each other.
% Signs of the loading variables are useful to compare within a direction vector; but these vectors
% can be rotated by 180°and still have the same interpretation.
%
%The contribution of all terms to each direction (PC) can also be extracted. 

%% Neu rausgekuerzt: 
%Regarding the first principle components, all terms have nearly the same weight. Being more precisely, term 5 has the least and term 3 the highest PC 1 value. Term 5 has the highest contribution to the second principle component, term 6 and term 3 have an very low weight that can be neglected. Regarding the third principle component, we recognize high values of term 6 followed by term 3; terms 1, 2, 4 and 5 have similar moderate weight. 

\begin{figure}[t]
    \centering
    \includegraphics[scale=.3]{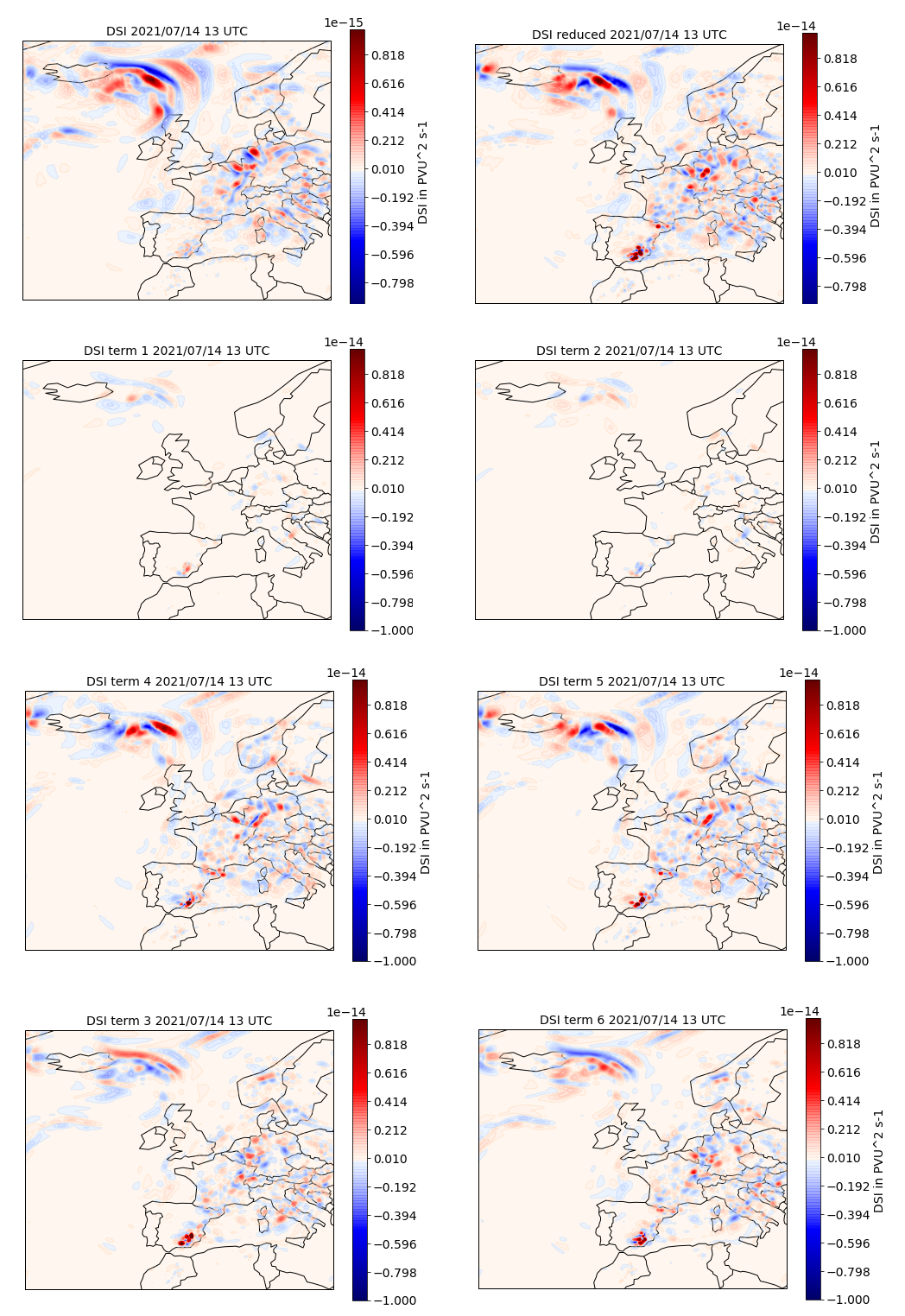}%%
\caption{\label{fig:DSI-terms}\small{The DSI (first row, first column) and its six terms (second-fourth row) are shown. 
The suggested reduced DSI (first row, second column) is given by the sum of terms 2, 5 and 6 (rows 2,3,4, second column). For the corresponding satellite image see the appendix.}}
\end{figure}

%Recalling that the principal components capture the most variation in a data set
The resulting loading plot with the PCs as axes is shown in Fig. \ref{fig:Loadings}.
 %that is a plot of the direction vectors that define the model. Here, the vectors are given by the blue arrows. 
 The plot depicts the contribution of each term to the PCs and compares the correlation with each other. The red point denotes the common origin.
Based on the diagonal orientation to each other term 1 and term 2 again show a distinct negative correlation. The same holds for term 4 and 5 as well as for terms 3 and 6. These results are also shown in the covariance matrix, see Fig. \ref{fig:CovMat_large}. 
Even though term 1 and term 2 are directed oppositely, they can be seen as cluster, being negatively correlated. 
Recalling the definitions of the DSI terms given in \eqref{eq:terms1-3} and \eqref{eq:terms4-6}
we recognize the same partial PV derivatives in the terms that form a cluster.
Term 1 and 2 contain $\partial \Pi / \partial p$;
term 4 and term 5 contain the derivative $\partial \Pi / \partial x$, terms 3 as well as term 6 contain the derivative $\partial \Pi / \partial y$. 
Therefore, we conclude that the PV seems to have a major contribution to the DSI field.
The DSI and all its six terms are shown for a case study in Fig. \ref{fig:DSI-terms}. 
The DSI fields and each single term are illustrated for the July 14th 2021 over the Euro-Atlantic sector. 
For the corresponding satellite image see the appendix. The negative correlations of terms 1 and 2, 4 and 5, 3 and 6 are confirmed. 

Therefore, we find three clusters of two terms indicating that the DSI can be approximated to three terms: 
\begin{equation} \label{eq:DSIreduction}
\rm{DSI} \approx \underbrace{ -\frac{\partial \Pi}{\partial p} \cdot \frac{\partial \theta}{\partial x} \cdot \frac{\partial B}{\partial y }}_{\text{Term 2}}
 + \underbrace{ \frac{\partial \theta}{\partial p} \cdot \frac{\partial B}{\partial y } \cdot \frac{\partial \Pi}{\partial x}}_{\text{Term 5}} +
\underbrace{- \frac{\partial \theta}{\partial p} \cdot \frac{\partial B}{\partial x} \cdot \frac{\partial \Pi}{\partial y}}_{\text{Term 6} }
\end{equation}
We note that terms 1 and 2 contribute less to the DSI signal such that a further reduction neglecting term 2 in \eqref{eq:DSIreduction} would possible. We further note that the evaluation of different case studies might lead to different results.

%\textcolor{red}{Figures X and Y show exemplary the PV and the DSI field, confirming the strong relation of the PV to the DSI fields (muss noch ergaenzt werden) }. 
%In addition to changes in the vorticity field, the DSI detects changes in the Bernoulli-function which contains the kinetic energy and in the potential temperature as thermodynamic quantity. 

%\begin{equation} \label{eq:DSI_approx}
%\mathrm{DSI} \approx \underbrace{
%- \frac{\partial B}{\partial y }\cdot %\frac{\partial \theta}{\partial x}
%\cdot\frac{\partial \Pi}{\partial p} 
%}_{\text{Term 2}}
%+ \underbrace{\frac{\partial B}{\partial p} \cdot \frac{\partial \theta}{\partial  x} \cdot \frac{\partial \Pi}{\partial y}
%}_{\text{Term 3}}
%+\underbrace{
% \frac{\partial \theta}{\partial p} \cdot \frac{\partial B}{\partial y } \cdot \frac{\partial \Pi}{\partial x}
%}_{\text{Term 5}}
%\end{equation}
%
%neglecting term 1, term 4 and term 6. 

The last step of the PCA is to project the original data onto a new feature space with less dimensions. It is spanned by the principle components. 
Consider a $6\times3$ dimensional matrix denoted as $\mathbf{P}$, where the three columns are the principle components. Let $\mathbf{X}$ denote the data matrix of size $n\times 6$ with $n$ observations of the 6 terms that we simply called Term 1-Term 6. 
Then, the Matrix $\mathbf{Z}= \mathbf{X}\cdot \mathbf{P} $ holds the synthetic variables, called scores. 

%\begin{itemize}

%\item[PC 2] Term1, 2, 4 and 5 have a good to comparatively high weight. Only term 3 and 6 have an extraordinary low weight which is neglectable.
%\item[PC 3] While term3 and 6 have a good to high weight, the terms 1, 2, 4 and 5 share a similar weight which can be accounted as moderate.
%\end{itemize}

\section{Discussion} 

Regarding the $\DSI$ as Jasobi-determinant of a $3\times 3$ dimensional matrix, the $\DSI$ can be formulated as sum of six terms given in \eqref{eq:terms1-3} and \eqref{eq:terms4-6}. 
We asked the question, which of these terms contribute most to the DSI signal? Are there DSI terms that can be neglected to minimize the computing time? In order to answer these questions, we applied a Principle Component Analysis to the DSI, more precisely to the six DSI terms. 

%We find that three Principal Components can maintain a large amount of variance explained of 96,9\%. We further find that three of the six DSI terms could indeed be neglected in future studies. 

In a first approach a PCA was applied to all nine single gradients \eqref{eq:DSIgrads} defining the DSI, which did not show reasonable results. This underlines the importance of coupling the three constitutive quantities potential temperature, Bernoulli funtion and PV that all together define the DSI.

Applying a PCA to the six DSI terms of the Jacobi-determinant, we first find that three Principal Components can maintain a large amount (99.9\%) of variance explained. Second, evaluating the DSI terms in the feature space, we further find that the main information of the DSI is captured by three DSI terms, i.e. the additional three DSI terms can be neglected. It turns out that even thought the PV gradients have the largest contribution to the DSI signal, the combination of the potential temperature gradient and the gradient of the Bernoulli function seems to be essential, which confirms the definition of the DSI..

Furthermore, the reduction of terms does not depend on the spatial and temporal scales. We have presented the results the PCA for the ERA5 data set with a horizontal resolution of about 30 km and a temporal resolution of hour. The same results are achieved for the ERA-Interim data set with a resolution of about 80 km, respectively 6 h, and for the COSMO-REA6 reanalysis data set with a resolution of 6 km and 1 h. Therefore, we find a scale independent approximation of the DSI. 
%We note that here, the 600 hPa surface is used, and the results might differ vertically at another surface.

To summarize, a first performance of the Principle Component Analysis applied to the six terms of the Dynamic State Index shows that the transformation and reduction of dimensions is reasonable. Even though, the potential vorticity seems to be the dominant quantity, all three constitutive quantities defining the DSI have an important contribution on the DSI signal to diagnose atmospheric developments. 

\section{Acknowledgements}
 This research is formed in the summer term 2020 by the course Geostatistics at the Institute for Geography of the Freie Universti\"at Berlin. We further thank Peter N\'evir for his time to discuss the results. 

\bibliography{references}

\end{document}